\documentclass[a4paper]{jpconf}
\usepackage{graphicx}
\newcommand{\X}{X(3872)}
\newcommand{\DD}{D^0\bar{D}^0}
\begin{document}
\title{Long-distance structure of the $X(3872)$}

\author{F.-K. Guo$^1$, C. Hidalgo-Duque$^2$, J. Nieves$^2$, A.
  Ozpineci$^3$ and  M. Pav\'on Valderrama$^4$}
\address{$^1$ Helmholtz-Institut f\"ur Strahlen- und
             Kernphysik and Bethe Center for Theoretical Physics, \\
             Universit\"at Bonn,  D-53115 Bonn, Germany}
\address{$^2$ Instituto de F\'isica Corpuscular (IFIC),
             Centro Mixto CSIC-Universidad de Valencia,
             Institutos de Investigaci\'on de Paterna,
             Aptd. 22085, E-46071 Valencia, Spain}
\address{$^3$ Middle East Technical University - Department of Physics
TR-06531 Ankara, Turkey}
\address{$^4$ Institut de Physique Nucl\'eaire,
             Universit\'e Paris-Sud, IN2P3/CNRS,
             F-91406 Orsay Cedex, France}

\ead{jmnieves@ific.uv.es}

\begin{abstract}
We investigate heavy quark  symmetries
for heavy meson hadronic molecules, and explore the consequences of  assuming
the $X(3872)$ and $Z_b(10610)$ as an isoscalar $D\bar D^*$
and an isovector $B\bar B^*$ hadronic molecules, respectively.
The symmetry allows to predict new hadronic molecules, in particular
we find an isoscalar $1^{++}$ $B\bar B^*$ bound state
with a mass about $10580$~MeV and  the isovector
charmonium partners of the $Z_b(10610)$ and the $Z_b(10650)$
states. Next, we study the $\X \to \DD\pi^0$ three body decay. This decay mode is more sensitive to the long-distance   
structure of the $\X$ resonance than its  $J/\psi\pi\pi$ and $J/\psi3\pi$ 
decays, which are mainly controlled by the short distance part of the $\X$
molecular wave function.   We discuss the $\DD$ final state
interactions, which in some situations become 
quite important. Indeed in these cases, a precise measurement of this partial decay
width could  provide precise information on the interaction strength between 
the $D^{(*)}\bar D^{(*)}$ charm mesons. 
\end{abstract}

\section{Heavy quark symmetries and hidden charm meson molecules}
\medskip
Heavy hadron molecules are a type of exotic hadron theorized more than
thirty years ago~\cite{Voloshin:1976ap,DeRujula:1976qd}.  Their main
component is a pair of heavy hadrons instead of a quark--antiquark
pair. The experimental advances in heavy quarkonium spectroscopy have
identified several molecular candidates among the recently observed
$XYZ$ states. The most promising ones are the
$X(3872)$~\cite{Choi:2003ue} and the twin $Z_b(10610)$ and
$Z_b(10650)$ states, to be called $Z_b$ and $Z_b'$,
respectively~\cite{Belle:2011aa,Adachi:2012im}. Among many different
interpretations of the $\X$, the one assuming it to be\footnote{From
  now on, when we refer to $D^0 \bar D^{*0} , D^+ D^{*-}$, or in
  general $D \bar D^*$ we are actually referring to the combination of
  these states with their charge conjugate ones in order to form a
  state with well-defined C-parity.} a $\left(D\bar D^*-D^*\bar D
\right)/\sqrt{2}$, hadronic molecule with quantum numbers $J^{PC} =
1^{++}$~\cite{Aaij:2013zoa} is the most promising.  Heavy quark
symmetries deduced from QCD provide new insights into the hadron
spectrum. Thus, heavy quark spin symmetry (HQSS) implies that
molecular states may appear in HQSS multiplets. From heavy flavor
symmetry (HFS), we know that the interaction among heavy hadrons is
roughly independent on whether they contain a charm or a bottom
quark. Combining both HQSS and HFS, various partners of the $\X$ and
the isovector $Z_b'^s$ states can be
predicted~\cite{Guo:2009id,Bondar:2011ev,Voloshin:2011qa,Mehen:2011yh,
  Nieves:2011zz,Nieves:2012tt,HidalgoDuque:2012pq,Guo:2013sya}. Some
of these predictions from \cite{Guo:2013sya} are collected in
table~\ref{tab:predictions}.  Moreover,
owing to heavy antiquark-diquark symmetry, the doubly heavy baryons
have approximately the same light-quark structure as the heavy
antimesons. As a consequence and thanks to this approximate symmetry
 the existence of a heavy meson-antimeson molecules
implies the possibility of a partner
composed of a heavy meson and a doubly-heavy baryon (triply-heavy pentaquarks)~\cite{Guo:2013xga}.
Indeed, HQSS heavily constrains also the low-energy interactions among
heavy hadrons \cite{Bondar:2011ev, Mehen:2011yh, Nieves:2012tt,
  HidalgoDuque:2012pq, AlFiky:2005jd}. As long as the hadrons are not
too tightly bound, they will not probe the specific details of the
interaction binding them at short distances. Moreover, each of the
constituent heavy hadrons will be unable to see the internal structure
of the other heavy hadron. This separation of scales can be used to
formulate an effective field theory (EFT) description of hadronic
molecules~\cite{Mehen:2011yh,Nieves:2012tt} compatible with the
approximate nature of HQSS.  At leading order (LO) the EFT is
particularly simple and it only involves energy-independent contact
range interactions, since pion exchanges and coupled-channel effects
can be considered subleading~\cite{Nieves:2012tt}. Thus, at very low
energies, the interaction between a heavy and anti-heavy meson
($D^{(*)}\bar D^{(*)}$ or $B^{(*)}\bar B^{(*)}$) can be accurately
described just in terms of a contact-range potential. 
The LO Lagrangian respecting HQSS contains four independent
terms in the SU(3) flavor
limit~\cite{HidalgoDuque:2012pq}. These are determined
by the isoscalar $C_{0A}$ and $C_{0B}$ and the isovector $C_{1A}$ and
$C_{1B}$ low energy constants (LEC's). The contact interaction
potential is used as kernel of a two body elastic Lippmann-Swinger
equation (LSE). The LSE shows
an ill-defined ultraviolet (UV) behaviour, and it requires a
regularization and renormalization procedure (we employ a standard
Gaussian regulator as in~\cite{HidalgoDuque:2012pq,
  Guo:2013sya}). Bound states ($D^{(*)}\bar D^{(*)}$ or $B^{(*)}\bar
B^{(*)}$ molecules) correspond to poles of the $T$-matrix
below threshold on the real axis in the first Riemann sheet of the complex energy. 

If we assume that the $\X$  and the isovector $Z_b(10610)$ resonances are $\left( D\bar D^*-D^* \bar D\right)/\sqrt{2}$ and 
$\left(B\bar B^*+B^* \bar B\right)/\sqrt{2}$ bound states, respectively, and 
use the isospin breaking information of the decays of the $\X$ into 
$J/\psi\pi\pi$ and $J/\psi\pi\pi\pi$, we can determine three linear
combinations among the four LECs $C_{0A}$, $C_{0B}$, $C_{1A}$ and 
$C_{1B}$ with the help of HQSS and HFS~\cite{HidalgoDuque:2012pq,
  Guo:2013sya}. Note the complex isospin dynamics of the $\X$ state
implied by the experimental  ratio $\mathcal{B}_{X} = \Gamma
\left[X \to J / \Psi \,\rho\, \right]/\Gamma \left[X \to J / \Psi
  \,\omega\,\right] = 1.3 \pm 0.5$ \cite{Choi:2011fc}. The isospin properties of the
$\X$ molecule are mainly determined by its mass, which is only
few hundreds of keV below the $D^0\bar  D^{*0}$ threshold,  making 
relevant the around 8 MeV difference between the threshold of the neutral and 
of the charged ($D^+ D^{*-}$) channels.  Further details are discussed
in  \cite{HidalgoDuque:2012pq,Gamermann:2009uq}.

\begin{center}
\begin{table}[h]

\caption{\label{tab:predictions}Heavy meson--heavy meson combinations
  having the same contact term as the $X(3872)$ and $Z_b(10610)$, and
  the predictions of the masses, which are understood to
  correspond to bound states except if we write ``V'' in parenthesis
  for denoting a virtual state. $\dagger$: increasing the strength of the
  potential to account for the various uncertainties, the virtual pole
  evolves into a bound state. Masses are given  (MeV units) for two
  UV regulators. For further details see \cite{Guo:2013sya}.}
\medskip
\footnotesize\rm
\centering
\begin{tabular}{ c c c c c c}
       $I(J^{PC})$ & States & Thresholds & $M$ ($\Lambda=0.5$ GeV) &
       $M$ ($\Lambda=1$ GeV) & Measurements
       \\\hline
        $0(1^{++})$ & $\frac1{\sqrt{2}}(D\bar D^*-D^*\bar D)$ &
       3875.87 & 3871.68 (input) &  3871.68 (input) &
       $3871.68\pm0.17$~\cite{PDG}  \\
                        $0(2^{++})$ & $D^*\bar D^*$ &
       4017.3  & $4012^{+4}_{-5}$  &  $4012^{+5}_{-12}$ & ?\\
        $0(1^{++})$ & $\frac1{\sqrt{2}}(B\bar B^*-B^*\bar B)$ &
       10604.4 & $10580^{+9}_{-8}$ &  $10539^{+25}_{-27}$ & ?\\
                        $0(2^{++})$ & $B^*\bar B^*$ &
       10650.2 & $10626^{+8}_{-9}$ & $10584^{+25}_{-27}$ & ?\\
                        $0(2^{+})$ & $D^*B^*$ &
       7333.7 & $7322^{+6}_{-7}$ & $7308^{+16}_{-20}$ & ?\\ \hline
        $1(1^{+-})$ & $\frac1{\sqrt{2}}(B\bar B^*+B^*\bar B)$ &
        10604.4 & $10602.4 \pm 2.0$ (input) & $10602.4 \pm 2.0$ (input) &
       $10607.2\pm2.0$~\cite{Belle:2011aa} \\
        & & & & & $10597\pm9$~\cite{Adachi:2012cx}\\
                        $1(1^{+-})$ & $B^*\bar B^*$ &
  10650.2 & $10648.1 \pm 2.1 $ &  $10648.1 ^{+2.1}_{-2.5}$ & $10652.2\pm1.5$~\cite{Belle:2011aa} \\
        & & & & & $10649\pm12$~\cite{Adachi:2012cx}\\
                        $1(1^{+-})$ & $\frac1{\sqrt{2}}(D\bar D^*+D^*\bar D)$ &
       3875.87 & $3871^{+4}_{-12}$ (V) &  $3837_{-35}^{+17}$ (V) & $3899.0 \pm 3.6 \pm 4.9$~\cite{BESIII:2013} \\
        & & & & & $3894.5 \pm 6.6 \pm 4.5$~\cite{Liu:2013xoa}\\
                        $1(1^{+-})$ & $D^*\bar D^*$ &
       4017.3 & $4013^{+4}_{-11}$ (V) &  $3983_{-32}^{+17}$ (V) & ? \\
                        $1(1^{+})$ & $D^*B^*$ &
       7333.7 & $7333.6^{\dagger}_{-4.2}$ (V) & $7328^{+5}_{-14}$ (V) & ?\\
   \end{tabular}
\end{table}
\end{center}

\section{$\X \to \DD\pi^0$ decay}
\medskip
As it is discussed in \cite{Guo:2014hqa}, in the hadronic
molecular picture, the decay channels of the $\X$  with a
charmonium in the final state ($J/\psi\pi\pi$,
$J/\psi3\pi$, $J/\psi\gamma$ and $\psi'\gamma$) are mainly sensitive
to the short distance part of the $\X$ wave-function. This is because
the heavy quarks of the $D\bar D^*$
meson pair have to recombine to get the charmonium in the final state. The 
transition from the charm--anti-charm meson pair into the $J/\psi$
plus pions (or a photon), occurs at a distance much smaller than both
the size of the $\X$ as a hadronic molecule and  the range of
forces between the $D$ and $\bar D^*$ mesons. However, in the case of
the $\X \to \DD\pi^0$ decay, one  of the constituent hadrons ($D^0$)
is in the 
final state and the rest of the final particles are products of the decay of 
the other constituent hadron ($\bar D^{*0}$) of the $\X$
molecule. Thus, in this decay the relative
distance between the $D\bar D^*$ mesons can be as large as
allowed by the size of the $\X$ resonance, since the final state is
produced by the decay of the $\bar D^*$ meson instead of a
rescattering transition. Actually, it can be proved that within some 
approximations, the $d\Gamma/d|\vec{p}_{D^0}|$ distribution 
is related to the $\X$ wave-function $\Psi(\vec{p}_{D^0})$ \cite{Guo:2014hqa}.

\begin{figure}[h]
\begin{minipage}{37pc}
\includegraphics[width=37pc]{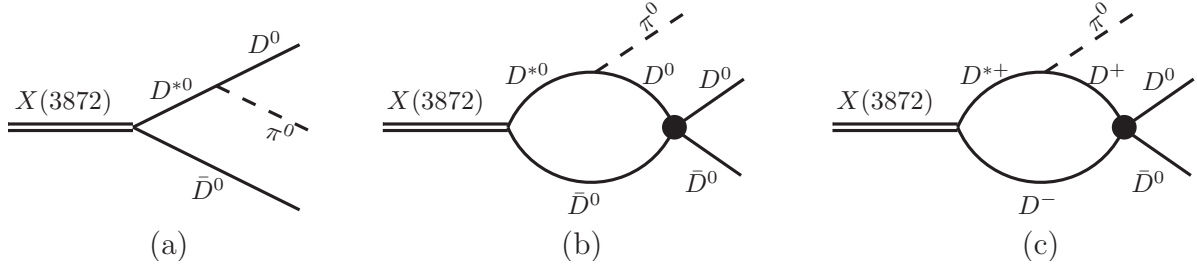}
\caption{\label{fig:fig1}Feynman diagrams for the decay
  $\X\to\DD\pi^0$. The charge conjugate
channel is not shown but included in the calculations.}
\end{minipage}
\end{figure}

\begin{figure}[h]
\begin{minipage}{37pc}
\includegraphics[width=17pc]{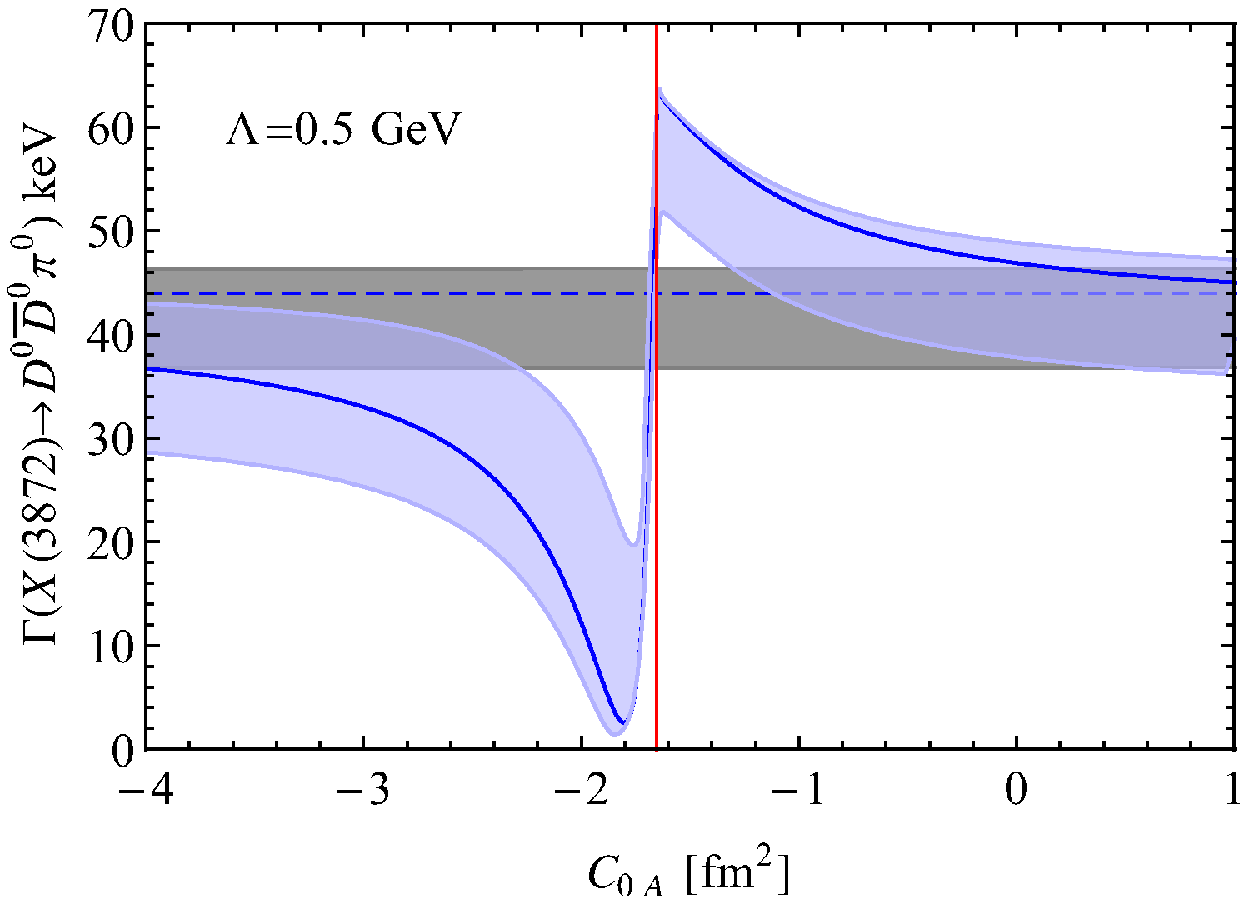}\hspace{3pc}\includegraphics[width=17pc]{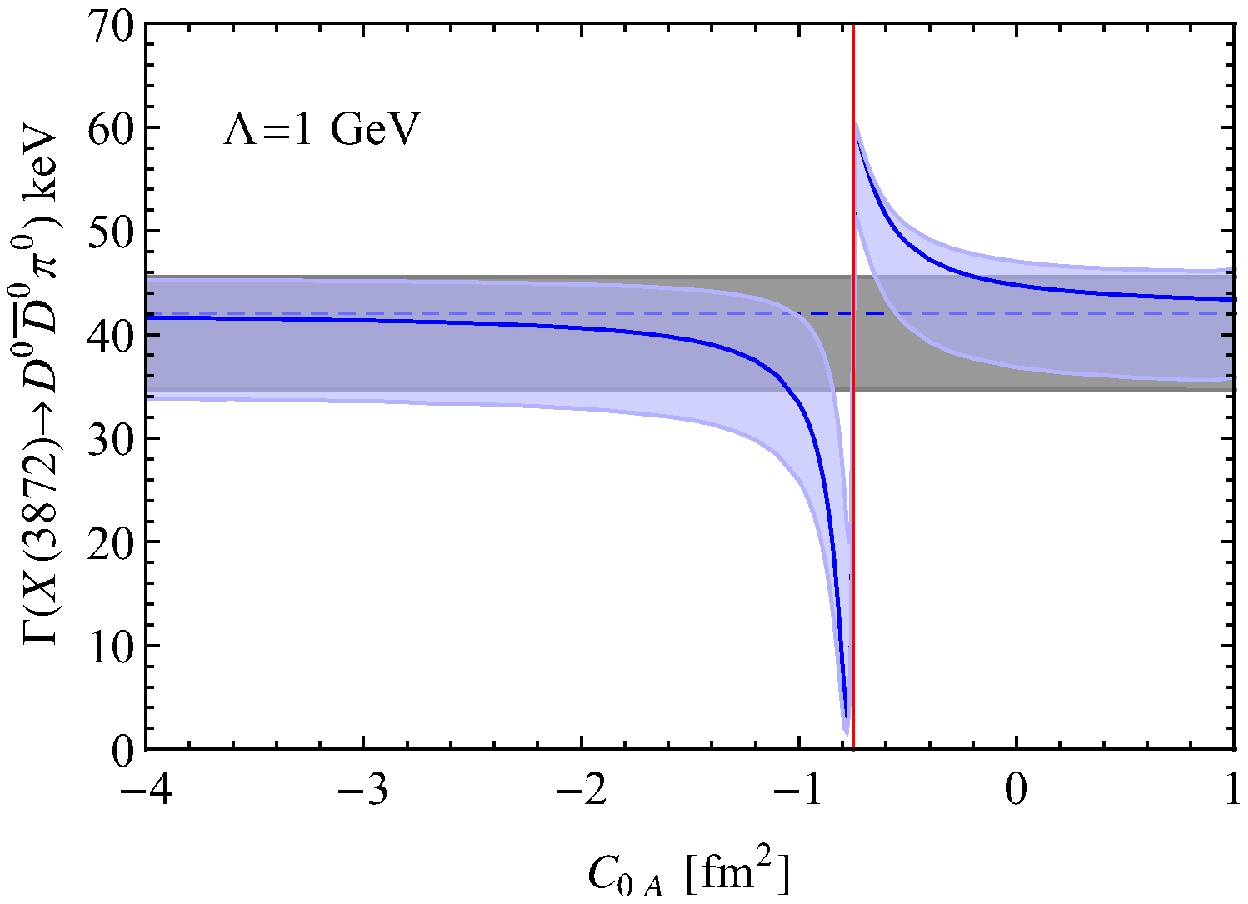}%
\caption{\label{fig:fig2}$\X\to D^0\bar D^0\pi^0$ partial decay
      width  as a function of  $C_{0A}$. The UV cutoff is set
      to $\Lambda=0.5$ GeV (1 GeV) in the left (right) panel.  The
      blue error bands contain $D\bar D$ FSI effects, while the grey
      bands stand for the tree level predictions (see Ref~\cite{Guo:2014hqa} for details).}
\end{minipage}
\end{figure}

To estimate the $\X \to \DD\pi^0$ decay width, we have evaluated the
diagrams depicted in figure \ref{fig:fig1}. The tree level
contribution corresponds to the mechanism depicted in the first
diagram of this figure. The amplitude is fully determined by the 
$D^0\bar D^{*0}\pi$ vertex,  the $\X$ mass and
its coupling constant to the neutral $D^0\bar D^{*0}$ channel, which is determined
by the residue of the $T-$matrix element at the $\X$ pole. We
find~\cite{Guo:2014hqa}  $\Gamma(\X\to D^0\bar D^0\pi^0)_{\rm tree} = 
44.0_{-7.2}^{+2.4}\left( 42.0_{-7.3}^{+3.6} \right)  ~{\rm keV}$, where the values outside and inside the parentheses are obtained
with UV cutoffs of $\Lambda=0.5$ and 1~GeV, respectively, and the uncertainty
reflects the uncertainty in the inputs ($M_{\X}$ and the ratio 
of decay amplitudes for the $\X\to J/\psi\rho$ and $\X\to
J/\psi\omega$ decays), and it is represented by the grey bands in
figure \ref{fig:fig2}. 

The last two diagrams in figure \ref{fig:fig1} account for the $ D \bar{D}\to
D\bar D$ final state interaction (FSI) effects, which are considered
by means of the appropriated linear combinations of the isoscalar and isovector $T-$matrices.
Those are obtained by solving a LSE in coupled channels with a LO
contact potential determined by the LEC's $C_{0A}$, $C_{0B}$, $C_{1A}$ and 
$C_{1B}$ introduced above~\cite{HidalgoDuque:2012pq, Guo:2014hqa}. As
mentioned,  the inputs (masses of the
$\X$ and $Z_b(10610)$ resonances and the ratio of $\X\to J/\psi\pi\pi$
and $\X \to J/\psi\pi\pi\pi$ branching
fractions) determine only  three of the four couplings, that
describe the heavy meson-antimeson $S$-wave interaction at LO in the
heavy quark expansion.  The value of the contact term
parameter $C_{0A}$ is undetermined, and thus 
the $D\bar D$ FSI effects on this decay are not fully determined.  As can be
seen in in figure \ref{fig:fig2}, these
effects might be quite large, because for a certain range of $C_{0A}$
values, a near-threshold isoscalar pole could be dynamically generated
in the $D\bar D$ system \cite{HidalgoDuque:2012pq, Nieves:2012tt}. If
the partial decay width is measured in future experiments, 
a significant  deviation from the predicted value at tree level  
will indicate a FSI  effect, which could eventually 
be used to extract the value of $C_{0A}$.  

\medskip
\ack
 C.~H.-D. thanks the support of the 
JAE-CSIC Program. This work is supported in 
part by the DFG and the NSFC through funds provided to the Sino-German CRC 110 
``Symmetries and the Emergence of Structure in QCD'', by the NSFC (Grant No. 
11165005), by the Spanish Ministerio de Econom\'\i a y Competitividad and 
European FEDER funds under the contract FIS2011-28853-C02-02 and the Spanish 
Consolider-Ingenio 2010 Programme CPAN (CSD2007-00042), by Generalitat 
Valenciana under contract PROMETEOII/2014/0068 and by the EU HadronPhysics3 
project, grant agreement no. 283286.

\end{document}